# Interface ferromagnetism and orbital reconstruction in BiFeO$_3$-La$_{0.7}$Sr$_{0.3}$MnO$_3$ heterostructures

P. Yu[1,¶,§], J.-S. Lee[2,¶], S. Okamoto[3], M. D. Rossell[4], M. Huijben[1,5], C.-H. Yang[1], Q. He[1], J.-X. Zhang[1], S. Y. Yang[1], M. J. Lee[1], Q. M. Ramasse[4], R. Erni[4], Y.-H. Chu[6], D. A. Arena[2], C.-C. Kao[2], L. W. Martin[7,8] and R. Ramesh[1,7]

[1] Department of Physics and Department of Materials Science and Engineering, University of California, Berkeley, CA 94720

[2] National Synchrotron Light Source, Brookhaven National Laboratory, Upton, NY 11973

[3] Materials Science and Technology Division, Oak Ridge National Laboratory, Oak Ridge, TN 37831

[4] National Center for Electron Microscopy, Lawrence Berkeley National Laboratory, Berkeley, CA 94720

[5] Faculty of Science and Technology, MSEA+ Institute for Nanotechnology, University of Twente, P. O. BOX 217, 7500 AE, Enschede, The Netherlands.

[6] Department of Materials Science and Engineering, National Chiao Tung University, Taiwan, 30010

[7] Materials Sciences Division, Lawrence Berkeley National Laboratory, Berkeley, CA 94720

[8] Department of Materials Science and Engineering, University of Illinois at Urbana-Champaign, Urbana, IL, 61801.

[¶] These authors contributed equally to this work.

[§] pyu@lbl.gov

**We report the formation of a novel ferromagnetic state in the antiferromagnet BiFeO$_3$ at the interface with La$_{0.7}$Sr$_{0.3}$MnO$_3$. Using x-ray magnetic circular dichroism at Mn and Fe $L_{2,3}$-edges, we discovered that the development of this ferromagnetic spin structure is strongly associated with the onset of a significant exchange bias. Our results demonstrate that the magnetic state is directly related with an electronic orbital reconstruction at the interface, which is supported by the linearly polarized x-ray absorption measurement at oxygen $K$-edge.**

The emergence of new states of matter at artificially constructed heterointerfaces is currently an intense area of condensed matter research.[1] Using transition metal perovskites as the building blocks, the charge[2], spin[3], and orbital[4] degrees of freedom can be controlled and manipulated at such interfaces. A significant amount of research is focused on the electronic reconstruction at the interface between LaAlO$_3$ and SrTiO$_3$.[2,5,6] The work of Chakhalian *et al.* demonstrated another novel aspect of interface control, namely the emergence of charge transfer driven orbital ordering and ferromagnetism in a (Y,Ca)Ba$_2$Cu$_3$O$_7$ (YBCO) layer at the interface with the doped manganite La$_{0.67}$Ca$_{0.33}$MnO$_3$ (LCMO).[4,7] Electric field control of such an interface ferromagnetic state would be a significant step towards magnetoelectric devices.[8,9,10,11,12] This begs the question: what would happen if the charge transfer at the interface is prohibited due to the ground state electronic structure of the transition metal species at the interface, such as the $d^5$ electronic state of $Fe^{3+}$.[13] We demonstrate in this paper that when such a $d^5$ system, for example manifested in the ferroelectric, antiferromagnet BiFeO$_3$ (BFO), is epitaxially conjoined at the interface to a multivalent transition metal ion such as $Mn^{3+}/Mn^{4+}$ in La$_{0.7}$Sr$_{0.3}$MnO$_3$ (LSMO), an unexpected ferromagnetic order is induced in the Fe sublattice at the interface as a consequence of a complex interplay between the orbital and spin degrees of freedom. The ferromagnetic state gives rise to a significant exchange bias interaction with the ferromagnetic LSMO, and both exhibit the same temperature dependence. The discovery of correlation between the electronic orbital structure at the interface and exchange bias suggests the possibility of using an electric field to control the magnetization of ferromagnet.

Heterostructures of the ferroelectric/antiferromagnet BFO (bottom layer, 30 nm) and ferromagnet LSMO (top layer, 5 nm) were grown on (001) Nb-doped STO substrates using pulsed laser deposition (see Supplementary information). X-ray absorption spectroscopy (XAS) spectra were acquired by recording the surface sensitive[7,14] total electron yield (TEY) current as a function of x-ray photon energy at beamline U4B of the National Synchrotron Light Source at Brookhaven National Laboratory. X-ray Magnetic Circular Dichroism (XMCD) was used to probe the ferromagnetic ordering of the heterostructure at the interface. 70% circular polarized x-rays were employed to gain higher beam intensities. Representative XAS and XMCD spectra taken at the Mn and Fe $L$-edges at 10 K are shown in Fig. 1a. The XMCD of ~ 23% at the Mn $L_3$-edge is consistent with previously measured values[15]. However, the ~ 4% XMCD observed at the Fe $L_3$-edge is surprisingly large considering that the nominal canted moment of BFO (~0.02-0.05 $\mu_B$/Fe)[16,17]. To rule out experimental artifacts, we carried out control measurement on a 30 nm BFO sample without the LSMO capping layer in which no measurable XMCD effect was observed (green data, Fig. 1b). Finally, we repeated the XMCD measurement on another LSMO/BFO heterostructure grown under identical conditions (Fig. 1b), which is essentially identical to the first dataset. This strongly suggests that in the few nanometers of the BFO film at the interface a new magnetic spin structure is present that is markedly different from that in the remainder of the BFO film. Additionally, the opposite signs of the Mn and Fe L-edges XMCD spectra (Fig. 1a) indicate that the coupling between the Mn and Fe across the interface is antiparallel. Although the actual spin structure at the interface could be complex, these XMCD spectra suggest that the coupling between the bulk LSMO spins and the bulk antiferromagnetic spin lattice of BFO is mediated through a very thin (a few unit cells) novel magnetic state localized at the interface.

To trace the origin of the magnetization at the interface, we have compared the dichroism of the interface BFO with reference samples, namely, ferrimagnetic $\gamma$-$Fe_2O_3$ and multiferroic $GaFeO_3$ [18] (Fig. 1b). The comparison between XMCD spectra for BFO and $\gamma$-$Fe_2O_3$ clearly shows that the dichroism of BFO is very different from that of $\gamma$-$Fe_2O_3$, specifically in that it lacks the reversal of the XMCD spectra corresponding to the $T_d$ site at

~710 eV. The comparison between the XMCD for BFO and GaFeO$_3$, on the other hand, reveals almost identical features, confirming the similarities in the lattice structures and electronic states of Fe in these two materials (O$_h$ site, Fe$^{3+}$). We can thus conclude that the relatively large magnetic moment in this heterostructure is arising from Fe$^{3+}$ ions on O$_h$ sites and unlikely to be the result of anion non-stoichiometry that might change the valence state of the Fe. Finally, electron energy loss measurements (Fig. S1c) across the interface confirm that the Fe$^{3+}$ is in the oxidation state.

The magnitude of the interfacial magnetization was quantitatively estimated using the XMCD spin sum-rule[19] (see supplementary information). We note that within the limits of the uncertainties in the sum rule estimation process for low 3d metals[20, 21], our calculated value ~ 3 $\mu_B$/Mn is consistent with our macroscopic SQUID measurement (Fig. 2a). The magnetization of the interfacial region of BFO layer is estimated to be ~0.6 $\mu_B$/Fe, which is surprisingly larger than the canted moment (0.03$\mu_B$/Fe) in the bulk BFO. Since the induced magnetization is confined at the interfacial region, the magnitude of the spin magnetization of BFO layer adjacent to LSMO is likely to be even larger than this value. The exact value of the moment is not a central finding in this paper; what is central is the significantly enhanced moment (localized near the interface) compared to the canted moment in the bulk and which is antiferromagnetically coupled to the LSMO bulk spins.

Macroscopic SQUID magnetometry was carried out at 10 K along the [100]$_{substrate}$ direction. Figure 2a shows a typical magnetic hysteresis loop for the heterostructure consisting of 5 nm LSMO and 30 nm BFO after field cooling from 350 K to 10 K in both +/- 0.2 T magnetic fields. We observe both a strong enhancement of the coercive field (~275 Oe) compared to that of the LSMO/STO sample (~40 Oe, red curve, Fig.2a) and a shift of the hysteresis loop opposite to the cooling field direction with an exchange bias (EB) field of 140 Oe.[22] Such an EB effect requires the presence of pinned, uncompensated spins in the antiferromagnet at the interface and is induced by the interface coupling between LSMO and BFO.[23, 24, 25, 26, 27]

Temperature dependent XMCD and SQUID measurements (Figure 2b) clearly demonstrate a strong interdependence between the ferromagnetic state in the Fe-sublattice at the interface and the exchange coupling between the two layers. As expected, the XMCD of Mn persists until ~ 300 K, consistent with the Curie temperature of the ultra-thin film LSMO.[28] The temperature dependent hysteresis loops show that the EB field vanishes at a blocking temperature ($T_B$) of ~100 K (this behavior was observed in all the 20 samples measured). An analogous behavior is found for the XMCD spectra of Fe (this was repeated in 3 samples), strongly suggesting that the novel magnetic moment in the Fe-sublattice is the source of the EB.

Similar temperature dependence is also observed in the linear dichroism at the oxygen *K*-edge (Fig. 3). The polarization directions of the linearly polarized x-rays (98% polarized) are tuned by rotating the x-ray incident angle, with 90° and 30° incident corresponding to complete in-plane (*E//a*) and majority of out-of-plane (*E//c*) polarized component, respectively (see the inset of Fig.3a). The linear dichroism signal could originate from both magnetic (spin)[29] and electronic (orbital) anisotropy[4, 30]. To exclude the contribution of the magnetic anisotropy from the induced magnetism, the data shown in the paper is taken without magnetic field; in addition, no change of the spectral shape or peak position was observed with an applied magnetic field (Fig. S5). Therefore, we emphasize that the observed linear dichroism in the current system is only related with the orbital anisotropy, while the contribution of the induced spins in the Fe sub-lattice is likely to be negligible.

Figure 3a shows the linearly-polarized XAS of the oxygen *K*-edge for the LSMO/BFO heterostructure. In contrast with the higher energy region (Bi-La-Sr *s*, *p* characters/Fe-Mn *s*, *p* characters), the lower energy region (O 2*p*-Mn (Fe) 3*d*) reflects the strong dependence on the linear polarization direction of incident x-rays. To investigate the hybridization process, the temperature dependence of the linearly-polarized XAS is shown in Figs.3b,c. Since the TEY signal comes from approximately the top 5-10 nm of the sample, the overall spectra are similar to that of the reference pure LSMO (top layer); however, the spectra also reveal information about the near-interface BFO. For example, the feature around

~530 eV (labeled as $P1_a$ and $P1_c$) corresponds to the mixture of Fe ($t_{2g}$ orbital) and Mn ($t_{2g}$ and $e_g$ orbitals) states, while the feature located at ~532 eV (labeled as $P2_a$ and $P2_c$) is related to only the $e_g$ levels of BFO (green curves in Figs. 3b,c). By following the features as a function of X-ray polarization direction (in-plane and out-of-plane), we obtain insight into the electronic orbital structure of the BFO at the interface. Figure 3d shows the temperature dependence of the peak positions of both the $P1$ and $P2$ features deduced from Fig. 3b,c. From 300 K down to 10 K, the $P1_a$ (red), $P1_c$ (blue) and $P2_a$ (green) features show a slight blue shift of the peak positions, due to a localization of the band at low temperature. On the other hand, a dramatic change for the spectra measured by out-of-plane polarized x-rays ($E//c$) is observed (see purple curve in Fig. 3d). The clear red shift of the peak position of the $P2_c$ feature ($d_{3z^2-r^2}$ orbital) suggests that normal to the interface a strong hybridized state between LSMO and BFO $d_{3z^2-r^2}$ orbitals via oxygen $2p$ orbital is formed below ~100 K. Again, the similar temperature dependence between the hybridization effect and the induced magnetization (as well as the EB effect) all point to a direct correlation between these observations and an electronic orbital reconstruction at the interface.

We now focus on bringing these experimental observations (Figs. 2 and 3) together to help explain the origin of ferromagnetism in the BFO layer at the interface as well as the resulting exchange bias coupling. Before taking into account the hybridization, the prerequisite of the Fermi energy continuity at the interface suggests the possible energy alignment as shown in Fig. 4a, in which the energy level of BFO is lower than that of LSMO due to the insulating nature of BFO and metallic nature of LSMO. However, the strong hybridization between the Mn and Fe $d_{3z^2-r^2}$ orbitals will form the bonding $d_{3z^2-r^2}$ orbital (lower energy) and antibonding orbital (higher energy) at the interface. While, the energy level of the $d_{x^2-y^2}$ orbitals in the Fe and Mn will not be significantly influenced due to the small coupling strength between them[31]. After the reconstruction, the electrons will take the lowest energy levels, i.e. bonding $d_{3z^2-r^2}$ orbital for the Fe site and $d_{x^2-y^2}$ orbital for the Mn site, which is confirmed by the oxygen $K$-edge study as shown in Fig.3. As a consequence, $d_{x^2-y^2}$ orbital ordering will be favored at the interface for LSMO. We note that in the rich phase diagram of manganites, orbital ordering is temperature dependent. For temperature above the transition temperature, $d_{x^2-y^2}$ orbital ordering would collapse. In this case, the forming of $d_{3z^2-r^2}$ antibonding orbital is energetically unfavorable and the system would reduce the bonding to lower the energy of antibonding orbital. Clearly, future studies would be necessary to testify this scenario and reveal the details of the coupling mechanisms.

We now turn to the magnetic coupling mechanism at the interface. From the Goodenough-Kanamori-Anderson (GKA) rules[32,33,34], the superexchange coupling between $Fe^{3+}$ and $Mn^{3+}$ (with $d_{x^2-y^2}$ orbital ordered in-plane, Fig.4b) and $Fe^{3+}$ and $Mn^{4+}$ are both expected to be strongly ferromagnetic. Moreover, the $d_{x^2-y^2}$ orbital ordering naturally leads to the antiferromagnetic coupling between the interfacial Mn layer and its neighboring Mn layer via the superexchange interaction between neighboring $t_{2g}$ spin and oxygen $2p$ orbital, which is responsible for the *A*-type (planar) antiferromagnetic ordering in metallic manganites such as $Pr_{1/2}Sr_{1/2}MnO_3$ and $Nd_{1/2}Sr_{1/2}MnO_3$[35]. Similar result has been reported with the *ab initio* calculations of the YBCO and LCMO interface.[36] Fig. 4b leads to the conclusion that the interfacial Fe spins and the Mn spins in bulk LSMO region are coupled antiferromagnetically, as is experimentally observed, in Figs. 1 and 2.

Having established this, we now examine the spin structure, in the BFO near the interface. Fig. 4c schematically describes the spin structure at the interface in the LSMO and the BFO layers. The competition between the ferromagnetic coupling across the interface triggered by the orbital order ( Fig. 4a,b), and the antiferromagnetic ground state of bulk BFO leads to a frustrated spin state with a large canting angle, Fig.4c. The magnitude of the canting angle is thus directly controlled by the strength of the interface coupling; the strong magnetic coupling between Fe and Mn at the interface due to the orbital ordering of LSMO would induce strong canting (magnetic moment) at the interface of BFO. We speculate that the EB effect is caused by the antiferromagnetic coupling between the interfacial Mn and the second Mn layers together with the induced moment in BFO. However, the induced moment in BFO must be pinned by an additional mechanism such as the spin anisotropy[26] in BFO or the interface roughness[25], which may cause a complicated magnetic domain structure. Clarifying such a microscopic structure is an important future direction.

To summarize, we have shown that at the interface between LSMO and BFO a new magnetic phase has been induced as a consequence of the electronic orbital reconstruction. It directly influences coupling between the BFO and the LSMO upon freezing at low temperatures and produces the pinned, uncompensated spins required for EB. Finally, we emphasize that changing the interface electronic state (doping level of Mn ions) by simply switching the polarization direction could in-principle modulate the interface magnetic coupling and eventually enable control of the magnetic state of the ferromagnet.[37]

Research at Berkeley was sponsored by the SRC NRI-WIN program as well as by the Director, Office of Science, Office of Basic Energy Sciences, Materials Sciences Division of the U.S. Department of Energy under contract No. DE-AC02-05CH1123. NSLS, Brookhaven National Laboratory, is supported by the U.S. DOE, Office of Science, Office of Basic Energy Sciences, under Contract No. DE-AC02-98CH10886. Work at ORNL was supported by the Division of Materials Sciences and Engineering, Office of Basic Energy Sciences, US Department of Energy. Y. H. C. also acknowledges the support of the National Science Council, R. O. C., under contract NSC 98-2119-M-009-016.

**Figures**

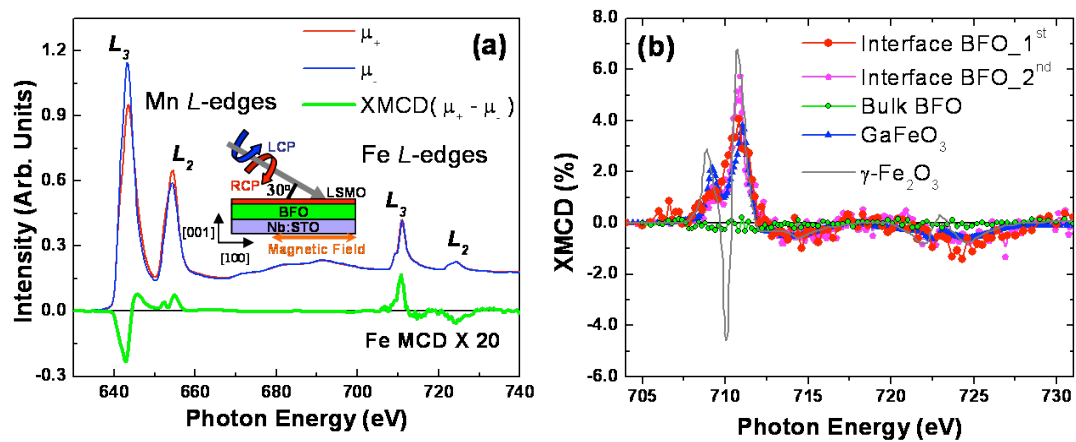

[Figure 1] (a) XAS and XMCD spectra of Mn and Fe $L_{2,3}$ edges taken at 10 K. The XMCD signal of Fe is multiplied by a factor of 20. (b) Comparison of the interface Fe XMCD with bulk BFO, GaFeO$_3$ and γ-Fe$_2$O$_3$. The spectra of GaFeO$_3$ and γ-Fe$_2$O$_3$ are normalized to the same scale as that of interface BFO state.

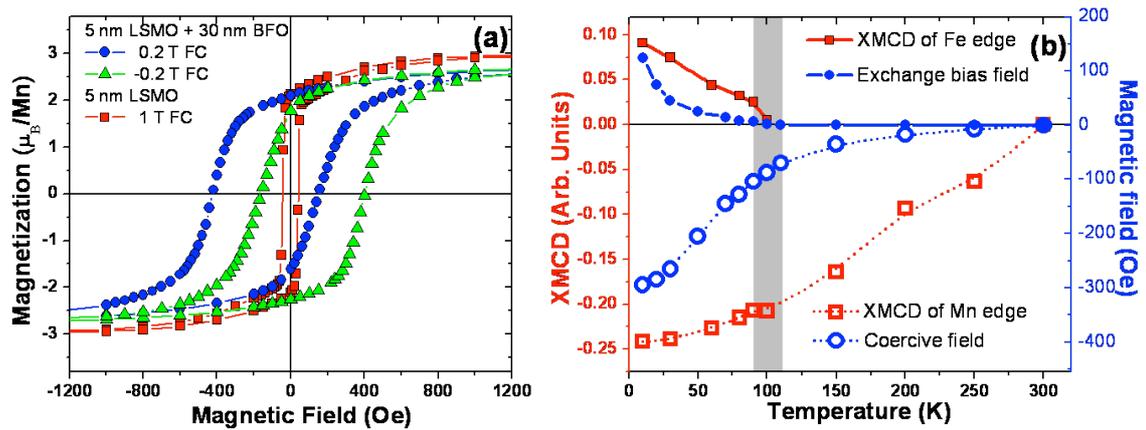

[Figure 2] (a) Magnetic hysteresis loops of a single LSMO layer and LSMO/BFO heterostructure measured along [100] direction at 10 K after +/- 0.2 T field cooling from 350 K, respectively. (b) Temperature dependence of the XMCD signal of Fe (solid red) and Mn (open red) compared with the exchange bias field (solid blue) and coercive field (open blue) as measured by SQUID.

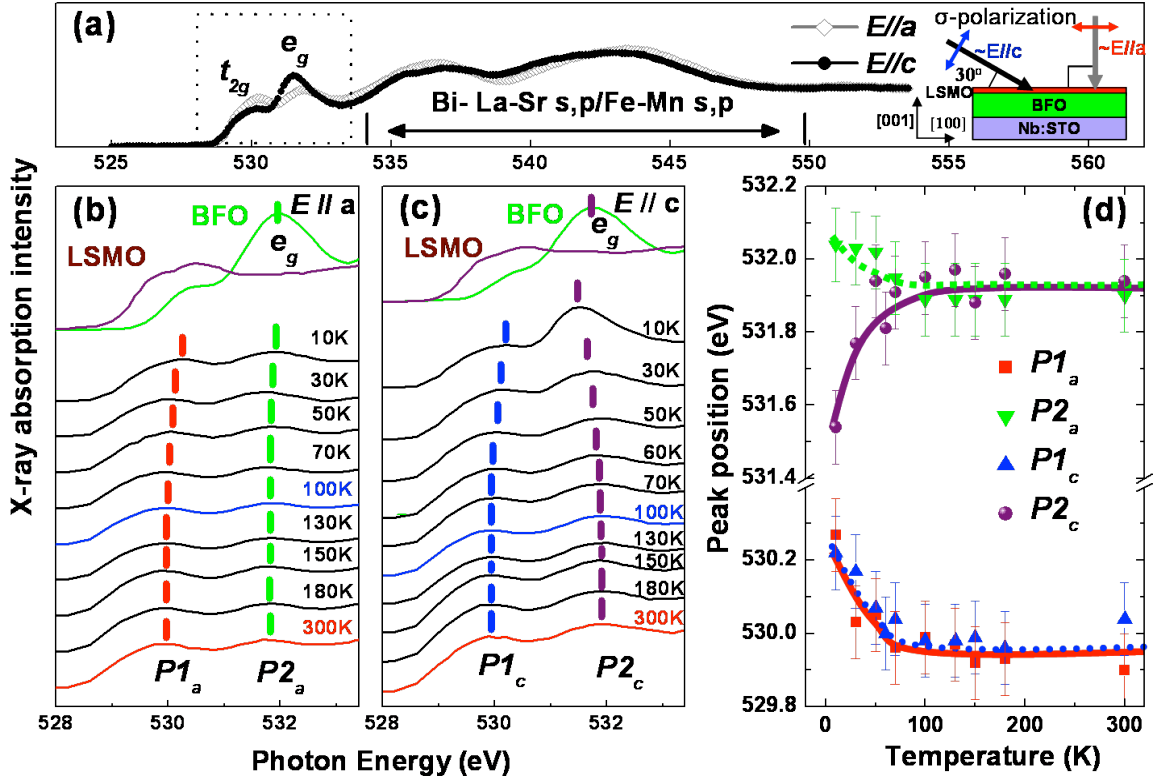

[Figure 3] (a) Polarization dependent oxygen *K*-edge XAS spectra with linearly polarized x-rays in wide energy range at *T*=10 K. Temperature dependent measurements of the polarized XAS spectra with polarization direction in-plane (b) and out-of-plane (c) for the specified bonding region of O 2*p*-Mn (Fe) 3*d*. (d) Temperature dependence of peak positions of interface orbitals. The red shift of $d_{3z^2-r^2}$ orbital indicates the strong hybridization between Mn and Fe across the interface.

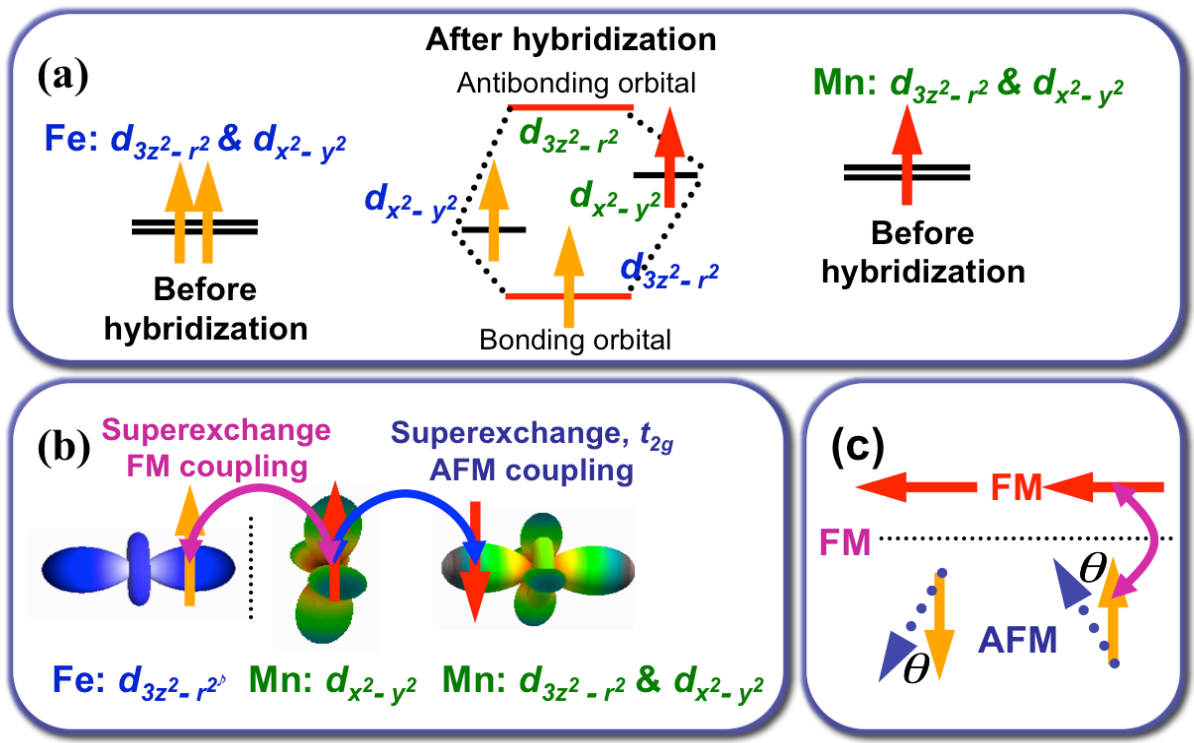

[Figure 4] (a) Schematic of the interface electronic orbital reconstruction, with hybridization. (b) Proposed interface spin configuration and coupling mechanism with $d_{x^2-y^2}$ orbital ordering in the interfacial LSMO. (c) Schematic of the origin of the interface magnetism.

**References**

1  J. Heber, Nature **459,** 28 (2009)
2  A. Ohtomo, and H. Y. Hwang, Nature **427**, 423 (2004).
3  K. Ueda, H. Tabata, and T. Kawai, Science **280**, 1064 (1998)
4  J. Chakhalian *et al.*, Science **318**, 1114 (2007)
5  N. Reyren *et al.*, Science **317,** 1196 (2007)
6  A. Brinkman *et al.*, Nat. Mater. **6**, 493 (2007)
7  J. Chakhalian *et al.*, Nature Physics **2**, 244 (2006)
8  M. Fiebig, J. Phys. D **38**, R123 (2005)
9  W. Eerenstein, N. D. Mathur, and J. F. Scott, Nature **442**, 759 (2006)
10 R. Ramesh and N. A. Spaldin, Nature Mater. **6**, 21 (2007)
11 S. W. Cheong and M. Mostovoy, Nature Mater. **6**, 13 (2007)
12 H. Béa *el al.*, J. Phys.: Condens. Matter. **20**, 433221 (2008)
13 C. H. Yang *et al.*, Nature Mater. **8**, 485 (2009)
14 H. Ohldag *et al.*, Phys. Rev. Lett. **87**, 247201 (2001)
15 J. J. Kavich *et al.*, Phys. Rev. B **76**, 014410 (2007)
16 C. Ederer and N. A. Spaldin, Phys. Rev. B **71**, 060401(R) (2005)
17 H. Bea *et al.*, Appl. Phys. Lett. **87**, 072508 (2005)
18 J. Y. Kim, T. Y. Koo and J. H. Park, Phys. Rev. Lett. **96**, 047205 (2006)
19 C. T. Chen *et al.*, Phys. Rev. Lett. **75,** 152 (1995)
20 Y. Teramura, A. Tanaka, and T. Jo, J. Phys. Soc. Japan 65, 1053 (1996)
21 A. Scherz, H. Wende and K. Baberschke, Appl. Phys. A **78**, 843 (2004)
22 W. H. Meiklejohn and C. P. Bean, Phys. Rev. **105**, 904 (1957)
23 J. Nogués and I. K. Schuller, J. Magn. Magn. Mater. **192**, 203 (1999).
24 H. Ohldag *et al.*, Phys. Rev. Lett. **91**, 017203 (2003)
25 A. P. Malozemoff, Phys. Rev. B 35, 3679 (1987)
26 N. C. Koon Phys. Rev. Lett. 78, 4865 (1997)
27 P. Miltenyi *et al.,* Phys. Rev. Lett. **84**, 4224 (2000)
28 M. Huijben *et al.*, Phys. Rev. B **78,** 094413 (2008)
29 G. Gan der Laan *et al.*, Phys. Rev. B **77**, 064407 (2008)
30 M. W. Haverkort *et al.,* Phys. Rev. B 69, 020408 (R) (2004)
31 Y. Tokura and N. Nagaosa, Science **288,** 462 (2000)
32 P. W. Anderson, Phys. Rev. **79**, 350 (1950)
33 J. B. Goodenough, Phys. Rev. **100**, 546 (1955)
34 J. Kanamori, J. Phys. Chem. Solid **10**, 87 (1959)
35 H. Kawano-Furukawa *et al.*, Phys. Rev. B **67**, 174422 (2003)
36 W. Luo *et al.,* Phys. Rev. Lett. **101**, 247204 (2008)

[37] S. M. Wu *et al.*, in preparation.